# Parameter Selection in Periodic Nonuniform Sampling of Multiband Signals

Moslem Rashidi, Sara Mansouri
Dep. of Signal Systems, Chalmers University of Technology, Goteborg, Sweden
e-mail: *moslem@student.chalemrs.se,masara@student.chalmers.se*

*Abstract*—Periodic nonuniform sampling has been considered in literature as an effective approach to reduce the sampling rate far below the Nyquist rate for sparse spectrum multiband signals. In the presence of non-ideality the sampling parameters play an important role on the quality of reconstructed signal. Also the average sampling ratio is directly dependent on the sampling parameters that they should be chosen for a minimum rate and complexity. In this paper we consider the effect of sampling parameters on the reconstruction error and the sampling ratio and suggest feasible approaches for achieving an optimal sampling and reconstruction.

## I. Introduction

Periodic nonuniform sampling is proposed for sampling of multiband signals and described in the articles [1]-[6]. It has been shown that for sparse spectrum signals the sampling rate can be reduced much lower than Nyquist rate, while retain sufficient information. The average sample ratio is specified by sampling parameters that can be approached to Landau's lower bound with proper selection of sampling parameters. However, the reduced sampling rates afforded by nonuniform scheme can be accompanied by increased error sensitivity. This paper is going to consider the sampling parameters and their effect on the sample ratio and reconstructed error.

The outline of the paper is as follows. In section II we provide a review of the sampling model and related definitions. Section III discusses the selection of sampling parameters $L$ and $p$ and finding the spectral index set from the band locations. Section IV shows the effect of sample pattern on the reconstructed signal in the presence of noise and provides a feasible algorithm to select a suitable sample pattern. A summary is given in section V.

## II. Sampling Model

We consider the class of continuous complex-valued multi-band signals of finite energy and maximum frequency of $f_{max}$, with band locations that are specified by a subset $\boldsymbol{F} = \bigcup_{i=1}^{N}[a_i, b_i)$ of the real line [4][6]. Assume the signal $x(t)$ of this class, is sampled nonuniformly at times $t = (nL + c_i)T, n \in \mathbb{Z}, 1 \leq i \leq p$, the samples then are categorized into $p$ sequences such that [4],[5]

$$x_i[n] = \begin{cases} x[nT], & n = mL + c_i, m \in \mathbb{Z} \\ 0, & otherwise \end{cases} \quad (1)$$

where, $T$ is the base sample time, $L$ is the period of pattern or block length, $p$ is the number of samples are kept in each block $L$ and $\boldsymbol{C} = \{c_1, c_2, \ldots c_p\}$ is the sample pattern [1].

The average sampling rate with choosing $T=1/f_{max}$ is [5]

$$D^- = \left(\frac{p}{L}\right) f_{max} \quad (2)$$

After taking *DFT* from both sides of (1) and represent in matrix form, the model of sampled signal in the frequency domain and in the interval $F_0 = [0, \frac{f_{max}}{L}]$ is given by [3]

$$\boldsymbol{y}(f) = \boldsymbol{A}\, \boldsymbol{z}(f), \quad \forall f \in F_0 \quad (3)$$

where $\boldsymbol{y}(f)$ is the known vector of observations as

$$\boldsymbol{y}(f)_{p\times 1} = \begin{bmatrix} X_1(f) \\ X_2(f) \\ \vdots \\ X_p(f) \end{bmatrix}, \quad \forall f \in F_0 \quad (4)$$

that $X_i(f)$ is the *DFT* of sequence $x_i[n]$, and $\boldsymbol{z}(f)$ is the unknown vector of signal spectrum as

$$\boldsymbol{z}(f)_{q\times 1} = \begin{bmatrix} X(f + \frac{k_1}{LT}) \\ X(f + \frac{k_2}{LT}) \\ \vdots \\ X(f + \frac{k_q}{LT}) \end{bmatrix}, \quad \forall f \in F_0 \quad (5)$$

that $X(f)$ is the Fourier transform of $x(t)$ and $\boldsymbol{A}$ is the known modulation matrix as

$$\boldsymbol{A}(i, l) = \frac{1}{LT} \exp\left(\frac{j2\pi c_i k_l}{L}\right), \quad 1 \leq i \leq p, 1 \leq l \leq q \quad (6)$$

The equation (3) relates the unknown spectrum of signal with the sampled data via modulation matrix $\boldsymbol{A}$.

If the spectrum of signal, $X(f)$, is sliced into $L$ slots and indexed from *0* to *L-1* , then the slots that are occupied by part of the signal, are called active slots and their indices collected into a set, $\mathbf{k}$, named spectral index set with length of $q$ such as [6],[3],[4]:

$$\mathbf{k} = \{k_1, k_2, \dots k_q\}, \quad q = |\mathbf{k}| \tag{7}$$

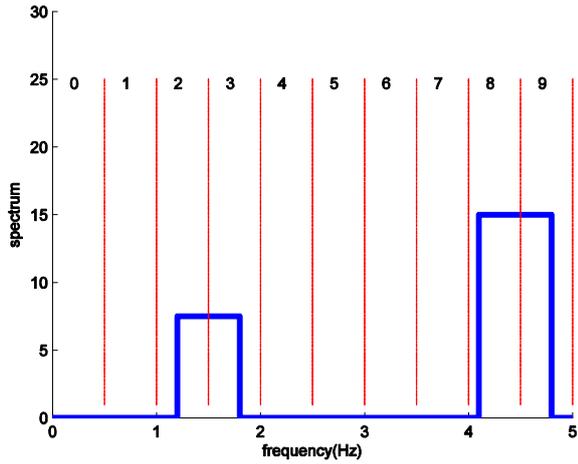

Fig. 1: Spectrum of multi-band signal is sliced into $L=10$ slots, the number of active slots is $q=4$ and the spectral index set is $\mathbf{k}=\{2,3,8,9\}$ [6],[3].

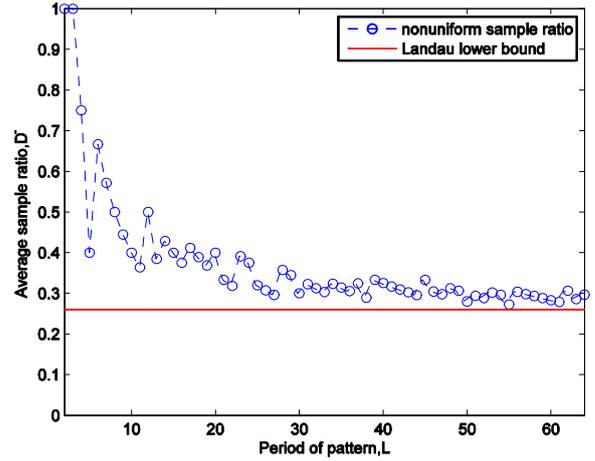

Fig. 2: Sample ratio versus the $L$ parameter, after $L=30$ the sample ratio becomes almost constant

Fig.1 shows the spectrum of a multi-band signal, sliced into $L=10$ slots, the number of active slots are $q=4$ and the spectral index set is $\mathbf{k}=\{2,3,8,9\}$.

The unique solution of (3) can be obtained using a left inverse-e.g. the pseudo-inverse of $\mathbf{A}$ as [3],[4]

$$\mathbf{z}(f) = \mathbf{A}^\dagger \mathbf{y}(f) \quad, \forall f \in F_0 \tag{8}$$

provided that $\mathbf{A}$ is full rank and $p \geq q$, hence the reconstruction can be obtained in time or frequency domain.

## III. SAMPLING PARAMETERS ($L, p$)

Selection of sampling parameters $L$ and $p$ can be considered in different aspects. In the sampling ratio sense from (2), it seems a large $L$ and small $p$ is desired for a minimal sample rate close to Landau's lower bound. On the other hand the required computations for reconstruction process is directly related to dimensions of $\mathbf{y}(f)$ and $\mathbf{A}$ that are determined by $L$ and $p$. Therefore, a tradeoff between sampling ratio and computation costs should be applied for selection of these parameters.

Since the parameter $p$ should be equal or bigger than $q$ for a unique solution of equation (3), first the value of $q$ for a given $L$ should be determined. This needs to find the spectral index set. Assume the band locations of the multiband signal, $x(t)$, is given in a subset $\mathbf{F} = \bigcup_{i=1}^{N}[a_i, b_i]$. With given $L$, the subset of active slots that are occupied by the $i$-th band $[a_i, b_i]$ is determined by

$$\left\lfloor \frac{a_i * L}{f_{max}} \right\rfloor \leq \mathbf{k}_i \leq \left\lfloor \frac{b_i * L}{f_{max}} \right\rfloor, \quad 1 \leq i \leq N \tag{9}$$

where $\lfloor \; \rfloor$ is the floor function. After finding subset $\mathbf{k}_i$ for all bands, the spectral index set is

$$\mathbf{k} = \bigcup_{i=1}^{N} \mathbf{k}_i \tag{10}$$

**Example**: Signal in Fig.1 with 2 bands that are located at $F=\{[1.2,1.8],[4.1,4.8]\}$, if $L=10$, $f_{max}=5$, the slots that are occupied by each band are respectively

$$\lfloor 1.2 * 10/5 \rfloor \leq \mathbf{k}_1 \leq \lfloor 1.8 * 10/5 \rfloor \Rightarrow 2 \leq \mathbf{k}_1 \leq 3$$
$$\Rightarrow \mathbf{k}_1 = \{2,3\}$$
$$\lfloor 4.1 * 10/5 \rfloor \leq \mathbf{k}_2 \leq \lfloor 4.8 * 10/5 \rfloor \Rightarrow 8 \leq \mathbf{k}_2 \leq 9$$
$$\Rightarrow \mathbf{k}_2 = \{8,9\}$$

hence $\mathbf{k} = \bigcup_{i=1}^{2} \mathbf{k}_i = \{2,3,8,9\}$ and $q=4$.

It is clear from (9) that increasing $L$ introduces more active slots and consequently needs more $p$ to compensate. In the other words we may choose a larger $L$ and still have the same or even worse average sample ratio. This can be seen in Fig.2 that depicts the average sample rate versus $L$ for the multiband signal of Fig.1. Also the figure illustrates the sample ratio becomes almost constant and close to Landau's lower bound when $L$ reaches a sufficient large value (here $L=30$) and then increasing $L$ more than this pint costs us more in term of computation, while it gains no significant sampling rate reduction.

Therefore, the parameters $L$ and $p$ could be optimized based on an intuitive consideration of computations and achieving minimum sampling rate with minimum value of $L$ and $p$. Because small values of $L$ may often suffice, and larger $L$ increases the computation cost of the reconstruction of the signal, small to moderate values of $L$ (e.g., in the tens to hundreds) are of interest [4].

## IV. SAMPLE PATTERN $C$

### A. Problem Statement

The sample pattern is the selection of $p$ out of $L$ numbers in the interval of 0 to $L-1$. A naive way of this selection is a bunched pattern or alternatively a random pattern, but these patterns can be applicable only in the ideal situations. We will see that choosing an inappropriate sample pattern can destroy the output and increase the sensitivity to noise in the reconstruction process.

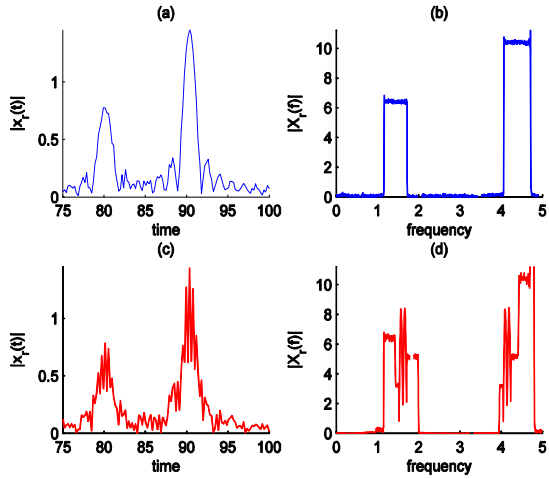

Fig. 3 Reconstructed signal from non-uniform samples in the presence of noise with two different sample patterns (a),(b) sequential pattern with cond(**A**)=1.3, relative error is 2.5% (c),(d) bunched pattern with cond(**A**)=128, relative error is 38%.

As mentioned before, $\mathbf{A}^\dagger$ exist if **A** is full rank also from (6) we see the rows of **A** are determined with sample pattern hence **C** should be such that making **A** full rank as the first criteria, this sample pattern then is called universal [1]-[6].

On the other hand, in practice the observation vector **y**(*f*) will be perturbed owing to *x*(*t*) being imperfectly band-limited to **F** or quantization and phase noise. Denoting the perturbation of **y**(*f*) with Δ**y**, owing to linearity the solution vector **z**(*f*) will be perturbed with Δ**z** such that

$$\mathbf{y}(f) + \Delta\mathbf{y} = \mathbf{A}\,(\mathbf{z}(f) + \Delta\mathbf{z}) \;\Rightarrow\; \Delta\mathbf{z} = \mathbf{A}^\dagger\,\Delta\mathbf{y} \quad (11)$$

It has been shown that [7]

$$\frac{\|\Delta\mathbf{z}\|}{\|\mathbf{z}+\Delta\mathbf{z}\|} \le \|\mathbf{A}\|.\,\|\mathbf{A}^{-1}\|\,.\,\frac{\|\Delta\mathbf{y}\|}{\|\mathbf{y}\|} \quad (12)$$

The inequality shows the relative change in the norm of the observation vector **y**(*f*) can be amplified by as much as $\|\mathbf{A}\|.\,\|\mathbf{A}^{-1}\|$. This number $cond(\mathbf{A}) = \|\mathbf{A}\|.\,\|\mathbf{A}^{-1}\|$ is called the condition number of matrix **A**, that can quantify the accuracy of solution of (3) [7].

Fig.3 shows the time domain and frequency domain of the reconstructed signal of Fig.1 from a nonuniform sampling setup in the presence of noise with two different sample patterns. The first sample pattern obtained from the sequential algorithm (will be mentioned later) with a low condition number of equal 1.3 and the other one uses a bunched pattern with condition number of 128. It is seen the reconstructed signal is highly disturbed owing to high condition number of bunched pattern such that the relative reconstructed error is around 38%.

Therefore, a sampling pattern that results in a well-conditioned of **A** is highly desirable as the second criteria [7].

A system of equation is considered to be well-conditioned if a small change in the observations vector results in a small change in the solution vector [7].

An ideal sample pattern is defined to give the *cond*(**A**)=1 among all patterns that are universal (at a fixed resolution *L*) for target set of spectral supports [3]. However depend on the spectral index set, **k**, sometimes is not possible to achieve condition number of one, then a pattern with smallest *cond*(**A**) is desired. Such a sampling pattern can be found as the solution to the following minimization problem [3]:

$$\mathbf{C}_{opt} = \arg\min_{\mathbf{C}:|\mathbf{C}|=p} cond\,(\mathbf{A}) \quad (13)$$

where the symbol |**C**| gives the cardinality or length of the set **C**. Solution of (13) by the exhaustive search would require $\binom{L}{p}$ evaluations of the condition of **A**, which is feasible only for small values of *L* and *p*. For the signal of Fig.1 that *L*=10, *p*=4 and **k**={2,3,8,9}, with $\binom{10}{4}=210$ evaluations, the result in table I is achieved. While the optimal pattern has a low condition number, the worst pattern has a huge condition number that can explode the result.

TABLE I
VARIOUS PATTERNS AND THEIR CONDITION NUMBERS, *L*=10, *P*=4

| Type | C | Condition Number |
|---|---|---|
| best pattern | {1,2,6,7} | 1.3 |
| worst pattern | {0,4,6,8} | $6\times 10^{16}$ |
| SFS pattern | {0,1,5,6} | 1.3 |

Although, a random selection of *p* numbers out of *L* would obtain a suitable pattern but it is highly related to the distribution of condition number that analytically cannot be easily achieved. The distribution of *cond(***A***)* for a typical case is shown in Fig.4. It shows that the most of the patterns place in the region of low condition number but still it is possibility to take a pattern with a high condition number. For example the probability of taking a pattern with the condition number less than 5.5 in this case is calculated to be 0.29 that is a pretty low value to guarantee a successful random pattern selection.

Exhaustive search is infeasible for large values of *L* and *p*; also random selection is not guaranteed all the time then we are looking for other search strategies to mitigate the number of search and obtain a low condition number. An interesting property is that all subsets of a sample pattern *C* will have lower condition number than *C*. This inspires us to use a sequential selection of elements of *C* as follows.

*B. Sequential Algorithm*

Given the set $\mathbb{L}=\{0,1,\ldots,L-1\}$, the goal is to find the subset $\mathbf{C}=\{c_1,c_2,\ldots,c_p\}$, with $p<L$ that minimizes the objective function *cond*(**A**). The sequential search starts from the empty set and sequentially adds the element $c^+$

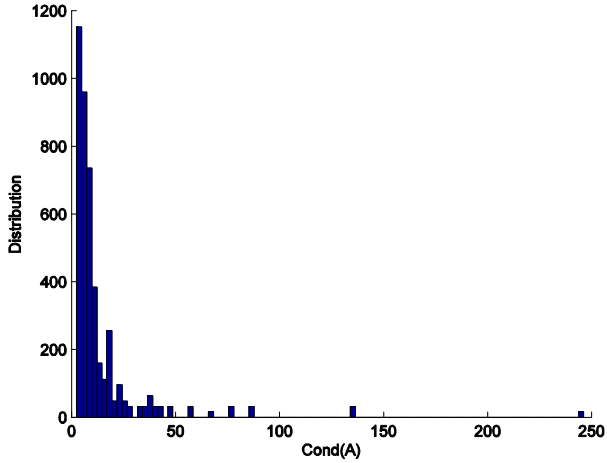

Fig.4: A typical distribution of condition number of **A**, sample patterns are generated randomly. Note, large values of condition number may happen.

that results in the minimum objective function when combined with the set $C_i$ that have already been selected. The search space is drawn like an ellipse to emphasize the fact that there are fewer states towards the full or empty sets [8]. The algorithm for choosing a sample pattern with sequential forward selection (SFS) is summarized below:

**SFS-Algorithm**

**Input**: *T, L, p, **k***

**Output:** sample pattern, **C**

1: Start with the empty set $\mathbf{C}_0 = \{\emptyset\}$

2: Select the next best element such that
$$c^+ = arg \min_{c \notin \mathbf{C}_i} [cond(\mathbf{A})]$$

3: Update    $\mathbf{C}_{i+1} = \mathbf{C}_i \cup c^+$; *i=i+1*

4: Go to step 2 if *i < p*

5: return  $\mathbf{C} = \mathbf{C}_p$

The algorithm is easy to implement and much faster than exhaustive search. Total number of search for choosing *p* number out of *L* in this way is derived with arithmetic progression as below:

The number of comparisons for selection of first element: *L*
The number of comparisons for selection of second element: *L-1*
…
The number of comparisons for selection of *p-th* element: *L-p+1*
Then total number of comparisons for the arithmetic progression is

$$S_p = pL - \frac{p(p-1)}{2} \qquad (14)$$

As an example when *L*=32 and *p*=10, an exhaustive search needs $\binom{32}{10} = 64 \times 10^6$ and with SFS search only 275 evaluations is needed. It shows a huge reduction in computations.

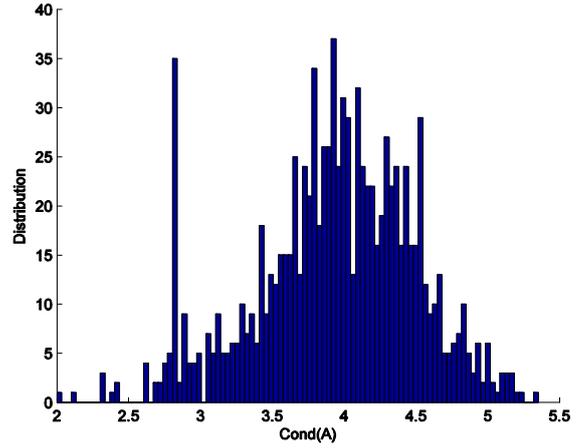

Fig.5: A typical distribution of condition number of **A**, sample patterns generated from the SFS-algorithm. Note, maximum condition number is 5.5.

For evaluation of the SFS algorithm, we randomly generate M=1000 different spectral supports and find the corresponding SFS patterns and condition numbers and plot the histogram in Fig.5. The result shows that the sample pattern achieved from the SFS-algorithm has a low condition number or sometimes the best one. Table I shows a SFS search result that the sample pattern gets the same condition number as the optimal one.

## V. SUMMARY

A sampling scheme named periodic nonuniform sampling for reducing the sampling rate of multiband sparse spectrum signals is revisited. The role of parameters on the sampling ratio and computation cost are considered. Also the concept of condition number and its effect on the reconstructed signal in presence of noise is described. In addition a feasible algorithm is presented to choose a suitable sample pattern.